\documentclass{elsart5}

\usepackage{graphicx}

\usepackage{amssymb}

\def\urs{URu$_{2}$Si$_{2}$}
\def\tord{$T_{\rm o}$}
\def\tsc{$T_{\rm c}$}
\def\tso{$T_{\rm c}'$}
\def\taf{$T_{\rm N}$}
\def\mord{${\mu}_{\rm o}$}

\def\mbohr{${\mu}_{\rm B}$}

\begin{document}

\begin{frontmatter}

\title{Pressure-temperature phase diagram of the heavy-electron superconductor URu$_{2}$Si$_{2}$}

\author[aff1]{H. Amitsuka\corauthref{cor1}}
\ead{amiami@phys.sci.hokudai.ac.jp}
\corauth[cor1]{}
\author[aff1]{K. Matsuda}
\author[aff1]{I. Kawasaki}
\author[aff1]{K. Tenya}
\author[aff2]{M. Yokoyama}
\author[aff3]{C. Sekine}
\author[aff4]{N. Tateiwa}
\author[aff5]{T.C. Kobayashi}
\author[aff6]{S. Kawarazaki}
\author[aff7]{H. Yoshizawa}
\address[aff1]{Department of Physics, Hokkaido University, Sapporo 060-0810, Japan}
\address[aff2]{Faculty of Science, Ibaraki University, Mito 310-8512, Japan}
\address[aff3]{Faculty of Engineering, Muroran Institute of Technology, Muroran 050-8585, Japan}
\address[aff4]{Advanced Science Research Center, Japan Atomic Energy Agency, Tokai 319-1195, Japan}
\address[aff5]{Department of Physics, Okayama University, Okayama 700-8530, Japan}
\address[aff6]{Department of Earth and Space Science, Osaka University, Toyonaka 560-0043, Japan}
\address[aff7]{Neutron Science Laboratory, ISSP, The University of Tokyo, Tokai 319-1106, Japan}
\received{10 September 2006}


\begin{abstract}
The pressure-temperature phase diagram of the heavy-electron superconductor {\urs} has been reinvestigated by ac-susceptibility and elastic neutron-scattering (NS) measurements performed on a small single-crystalline rod (2 mm in diameter, 6 mm in length) in a Cu-Be clamp-type high-pressure cell ($P <$ 1.1 GPa). At ambient pressure, this sample shows the weakest antiferromagnetic (AF) Bragg reflections reported so far, corresponding to the volume-averaged staggered moment of {\mord}$\sim$ 0.011 ${\mu}_{\rm B}$/U. Under applied pressure, the AF scattering intensity exhibits a sharp increase at $P \sim 0.7$ GPa at low temperatures. The saturation value of the AF scattering intensity above 0.7 GPa corresponds to {\mord} $\sim$ 0.41 ${\mu}_{\rm B}$/U, which is in good agreement with that ($\sim 0.39 {\mu}_{\rm B}$/U) observed above 1.5 GPa in our previous NS measurements. The superconductivity is dramatically suppressed by the evolution of AF phase, indicating that the superconducting state coexists only with the hidden order phase. The presence of parasitic ferro- and/or antiferromagnetic phases with transition temperatures $T_{1}^{\star} =$ 120(5) K, $T_{2}^{\star} =$ 36(3) K and $T_{3}^{\star} =$ 16.5(5) K and their relationship to the low-$T$ ordered phases are also discussed. 
\end{abstract}

\begin{keyword}
\PACS 74.70.Tx \sep 75.20.Hr \sep 75.50.Ee
\KEY {\urs} \sep hidden order \sep superconductivity \sep antiferromagnetism \sep hydrostatic pressure
\end{keyword}

\end{frontmatter}


\section{Introduction}\label{sec1}

The relationship between superconductivity and magnetism has been a central issue in strongly correlated electron systems including heavy-electron (HE), cuprate, ruthenate and organic superconductors. The superconducting (SC) state of these systems develops, to varying extents, in the proximity of a magnetically ordered phase, and thus magnetic fluctuations have been believed to mediate (or at least play an important role in) the Cooper pairing. One of the characteristic features seen in the HE superconductors is perhaps that the SC state occurs clearly under magnetically (or some other) ordered phases. Here, f electrons take part in both the orders. Limiting to pure (undoped) compounds and ambient pressure, such cases have been discussed in the antiferromagnets with relatively large staggered moments (UPd$_{2}$Al$_{3}$ \cite{geibel91}, UNi$_{2}$Al$_{3}$ \cite{geibel91_2}, CePt$_{3}$Si \cite{bauer04}), a very weak (probably short-range-ordered) antiferromagnet (UPt$_{3}$ \cite{aeppli88}), a ferromagnet (URhGe \cite{aoki01}) and a so-called hidden order (HO) compound ({\urs} \cite{broholm87}). In the present paper, we focus our attention on {\urs}, and report on our latest high-pressure studies. The experimental results obtained demonstrate that the superconductivity of this compound coexists with HO, but not with the AF order induced by applying pressure. We also show that the weak antiferromagnetism and ferromagnetism, both observed around HO, are extrinsic to the HO and SC states.  

{\urs} is crystallized in the ThCr$_{2}$Si$_{2}$ type structure with space group I4/mmm. The HO mentioned above occurs at {\tord} $\sim$ 17.5 K, where a portion of the Fermi surface disappears due to a gap opening, which is indicated by, for example, a reduction of the linear specific-heat coefficient from $C/T \sim 180$ mJ/K$^{2}$mol at $T \sim$ {\tord} to $\sim 60$ mJ/K$^{2}$mol at $\sim 2$ K \cite{palstra85,schlabitz86,maple86}. In 1987, Broholm et al. found from elastic neutron-scattering (NS) experiments that the 5f magnetic moments of U ions order at around {\tord}, with a propagation vector $Q = (1, 0, 0)$ \cite{broholm87}. The moments align along the $c$ direction with a saturation value of {\mord} $\sim$ 0.03 {\mbohr}/U. The 5f electrons on the remnant Fermi surface further undergo the SC transition at $\sim$ 1.2 K. Therefore, in the early stage of investigation this system was regarded as a HE compound where superconductivity coexists with a weak AF ordering. However, the internal fields which should be generated by the AF order were not detected in $^{29}$Si-NMR \cite{kohara86} and ${\mu}$SR measurements \cite{maclaughlin88}. This might imply that the order is short range with fluctuating moments, but such an interpretation is inconsistent with the fact that a very sharp anomaly is seen in various macroscopic quantities at {\tord}, particularly the mean-field like behavior of $C(T)$ with a large entropy reduction of ${\Delta}S \sim 0.2R\ln{2}$. (This provides a remarkable contrast to the weak AF order of UPt$_{3}$, where the order is characterized by the moments of similar magnitude, but not accompanied by any macroscopic anomaly at its onset temperature $\sim$ 5 K.) 

To resolve this contradiction, so far a variety of theoretical models have been put forward. They can be classified into two groups depending on whether the order parameter (OP) is (A) a magnetic dipole $m$ \cite{ge87,sikkema96,okuno98,yamagami00,bernhoeft03,mineev05} or (B) some other degree of freedom ${\psi}$ \cite{miyako91,bar93,santini94,ami94,bar95,kasuya97,ikeda98,ohkawa99,chandra02,virosztek02,kiss05,varma06}. In the former group, the AF order is regarded as intrinsic and the models involve a mechanism for decreasing the effective $g$ value to account for the difference between {\mord} and ${\Delta}S$. On the other hand, the models of the group B assume the AF order to be basically unnecessary for freezing the primary OP. In this case, the AF order further admits of two different interpretations: (B1) {\it a secondary order} induced by the development of ${\langle}{\psi}{\rangle}$, and (B2) a {\it second phase} independent of (or competing with) the HO. It is thus very important to clarify the origin of the weak AF order, to identify the hidden OP.

\section{A brief review of previous high-pressure studies on {\urs}}\label{sec2}

In 1999, we performed elastic and inelastic NS measurements on {\urs} under hydrostatic pressure, and observed that the AF Bragg scattering intensity $I_{\rm AF}$ is significantly enlarged by roughly 100 times with increasing pressure \cite{ami99}. The magnetic structure is unchanged, and {\mord} estimated reaches $\sim$ 0.39 {\mbohr}/U at 1.5 K for $P \sim$ 2 GPa. The NS measurements, however, provide magnetic signals averaged over the crystal volume, i.e., {\mord} $= \sqrt{v_{\rm AF}{\mu}_{\rm AF}^{2}}$ $({\propto} \sqrt{I_{\rm AF}})$, where $v_{\rm AF}$ and ${\mu}_{\rm AF}$ represent the AF volume fraction and the true magnitude of the staggered moment, respectively. NMR can distinguish these two quantities, and Matsuda et al. established from their high-$P$ $^{29}$Si-NMR measurements that what varies with pressure is not ${\mu}_{\rm AF}$ but $v_{\rm AF}$ \cite{matsuda01}. Our succeeding zero-field ${\mu}$SR measurements also shown that the AF and the seemingly non-magnetic HO phases are distributed in separate regions of the crystal \cite{ami02,amato04}. 

Motoyama et al. provided the thermodynamical evidence for the presence of the $P$-induced phase transition by measuring the coefficients of linear thermal expansion on a {\urs} single crystal under hydrostatic pressure \cite{motoyama03}. They observed that the $c/a$ ratio exhibits a significant increase when the system enters into the high-$P$ AF phase. They also showed that the location of the phase boundary line between the HO and AF phases sensitively depends on the sample quality. 

In parallel, we performed the NS measurements under uniaxial stress ${\sigma}$, and observed that $I_{\rm AF}$ increases only when ${\sigma}$ is applied along the basal $c$ plane \cite{yokoyama03}. From the detailed strain analyses on both the $P$ and ${\sigma}$ effects, we proposed that the axial strain ${\Delta}\ln{(c/a)} \equiv \hat{\eta}$ is the key parameter to drive the HO-to-AF transition. The critical value of ${\eta}$ is estimated to be very small ($\hat{\eta} \sim 10^{-3}$), implying that the heterogeneously ordered state observed may be ascribed to a weak distribution of strain in a crystal. 

Recently, Bourdarot et al. also presented a $P$-$T$ phase diagram based on the high-$P$ NS measurements performed by using a hydrostatic helium cell as well as a clump-type pressure cell \cite{bourdarot04}. They determined the first-order phase transition temperature as a midpoint of the increase in $I_{\rm AF}(T)$, and suggest that the first-order phase boundary line may have a critical end point, not meeting the second order phase boundary line $T_{\rm o}(P)$. They consider the symmetry of OP to be unchanged by pressure, and call the low and high-$P$ phases the small-moment and large-moment AF phases, respectively. They also reported that the transition was sharpened by using the helium pressure cell.

Through these high-$P$ studies, one can safely say that {\urs} undergoes a first-order phase transition by applying $P$ and ${\sigma}$ at low temperature. However, at least the following three points remain unclear. First, there is an inevitable debate on whether the first-order phase boundary line terminates at a critical end point or a bicritical point: topology of the $P$-$T$ phase diagram is very important for analyzing the symmetry of OP as mentioned above \cite{mineev05}. Secondly, it is unclear how the SC state behaves under pressure. In an earlier experiment, it was reported that {\tsc} linearly decreases with increasing $P$, but still remains finite at $P \sim$ 1.2 GPa \cite{mcelfresh87}, where we now know the AF phase to be dominant. On the other hand, very recently Uemura et al. have presented a new $P$-$T$ phase diagram based on thermal expansion and ac-susceptibility measurements, and suggested that the superconductivity does not coexist with the high-$P$ AF phase \cite{uemura05}. The third question has also been raised by their new phase diagram: the nature of a weak ferromeganetic (FM) phase developing below $\sim$ 35 K. The phenomenon itself has been known for a long time, but they found that its onset temperature decreases with $P$ and shows a tendency to merge into the N\'{e}el temperature {\taf} above $\sim$ 1 GPa.

In short, those three issues require further investigation to obtain the correct $P$-$T$ phase diagram of this system. A major difficulty to do this arises from the strong dependence of the $P$-induced phenomena on the sample and pressure quality, which prevents us from comparing simply the experimental data taken for different samples and/or different pressure conditions. We now adopt the same combination of a sample and a pressure cell throughout our investigation to obtain a reliable phase diagram. 

\section{Experimental procedure}\label{sec3}

A cylindrical {\urs} single crystal was grown in a tetra-arc furnace under a pure argon atmosphere by the Czochralski method. The dimensions of the grown crystal were about 25 mm in length and 2 mm in diameter (the cylindrical axis was parallel to the $c$ axis). The crystal was cleaved along the $c$ plane to a 6-mm-long cylinder to fit a pressure cell used for the present study, and subsequently annealed for 7 days at 1000 $^{\circ}$C. No spark-erosion process was applied.

NS measurements were performed on the general purpose triple-axis spectrometer (GPTAS) installed at the beam port 4G of the JRR-3M reactor in Japan Atomic Energy Agency. Pyrolytic graphite PG(002) crystals were used as a monochromator and an analyzer. Elastic scans were made down to 1.5 K under several pressures up to 1.07 GPa, with a neutron wavelength ${\lambda}$ = 2.334 {\AA} and a 40'-40'-40'-80' horizontal collimation. Higher-order reflections were suppressed by using two 40-mm-thick PG filters.

Hydrostatic pressure was applied by using a Cu-Be clamp cell (length 25 mm, inner diameter 2.6 mm, outer diameter 10 mm) with Fluorinert 70 \& 77 pressure-transmitting medium. The applied pressure was monitored in situ by measuring the ac susceptibility on the superconducting transition of a small piece of lead, which was encapsulated together with the sample in the pressure cell. For the ac susceptibility measurements, the standard Hartshorn-bridge technique was used with an external ac field of $\sim$ 1 Oe (peak to peak) and 70 Hz. The coaxial coils were wounded on an aluminum bobbin surrounding the pressure cell.

The low-$T$ ac-susceptibility measurements for studying the superconductivity were performed in the $P-T$ range 0.12 $< T <$ 10 K and $P <$ 0.76 GPa, using the same {\urs} and lead with the same pressure cell. The amplitude and frequency of applied field are the same as those given above, and a bakerite coil bobbin was used to avoid the generation of induction heat.

The dc-magnetization measurements for studying FM anomalies were performed down to 2 K on different batches of {\urs} and Rh-doped crystals, using a commercial SQUID magnetometer (MPMS, Quantum Design Co.).

\section{Results and discussions}

Fig. 1 shows temperature dependence of the NS Bragg scattering intensity $I_{\rm AF}$ of the (100) magnetic reflection for {\urs}, observed at ambient pressure by ourselves and four other groups. This is so-called the weak or small-moment antiferromagnetism of this system. The difference seen in the temperature dependence as well as the saturation value $I_{\rm AF}(T \to 0)$ is striking. The single crystal prepared for the present study exhibits the smallest magnitude of $I_{\rm AF}(T)$. Moreover, its onset temperature is the lowest ($\sim$ 13.5 K) among the reports, and apparently lower than the HO transition temperature {\tord} = 17.6(1) K, which is defined as the midpoint of the $C(T)$ jump. In contrast to the reports of NS measurements, previous studies on the macroscopic properties have never indicated such strong sample dependence of the 17.5 K transition. These NS data thus strongly suggest that the weak AF order with $Q = (1, 0, 0)$ is not intrinsic to the HO.

  \begin{figure}[t]
\includegraphics[scale =1]{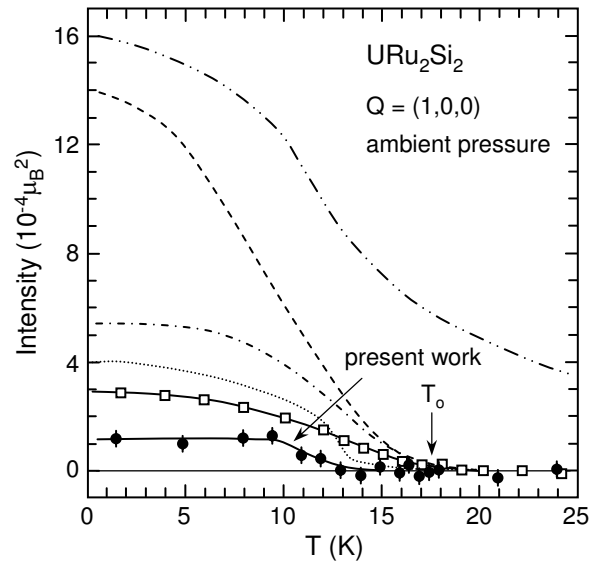}
\caption{
Neutron-scattering intensity of the magnetic Bragg peak of {\urs} at $Q = (1,0,0)$ as a function of temperature. The present results (closed circles) are compared with the previous experimental data reported by ourselves (open squares) \cite{ami99}, Broholm {\it et al.} (dash-double-dotted line) \cite{broholm87}, Mason {\it et al.} (broken line) \cite{mason}, F{\aa}k {\it et al.} (dash-dotted line) \cite{fak} and Honma {\it et al.} (dotted line)\cite{honma}.
}
  	\label{fig-1}
  \end{figure}

In Fig. 2, we show the $I_{\rm AF}(T)$ data taken for our new {\urs} single crystal at several pressure points up to 1.07 GPa. We observed that $I_{\rm AF}$ remains very small in the $P$ range below 0.6 GPa, and exhibits a dramatic increase between 0.6 and 0.8 GPa. The $I_{\rm AF}(P)$ curve at $T = 1.5$ K is characterized by a single-step increase at $P$ = 0.7 GPa, which provides a remarkable contrast to our previous NS data \cite{ami99} showing a nonmonotonous increase over a wide $P$ range (Fig. 3). Good agreement is seen in the saturation values of $I_{\rm AF}$ between the present and previous results, from which one can estimate ${\mu}_{\rm AF}$ to be 0.41 and 0.39 {\mbohr}/U, respectively. It can also be seen from Fig. 3 that the increase in the present $I_{\rm AF}(P)$ is much sharper than that in $v_{\rm AF}(P)$ obtained by NMR and ${\mu}$SR measurements \cite{matsuda01,ami02,amato04}, if plotted on a normalized scale. On the basis of NMR and ${\mu}$SR studies, the $I_{\rm AF}(P)$ curves at this low $T$ simply represent the pressure dependence of $v_{\rm AF}$. The observations thus indicate that the new sample has the narrowest $P$ range of the phase separation. We speculate that this has been brought about by the use of a small single crystal. It is expected that a smaller crystal will have a narrower strain distribution at ambient pressure, and also allow us to perform more homogeneous pressurization. In other previous studies shown in Fig. 1 and Fig. 3, the original single crystals have larger dimensions ($\gtrsim$ 4 mm in diameter) than the present crystal ($\sim$ 2 mm). 

  \begin{figure}[t]
\includegraphics[scale =1]{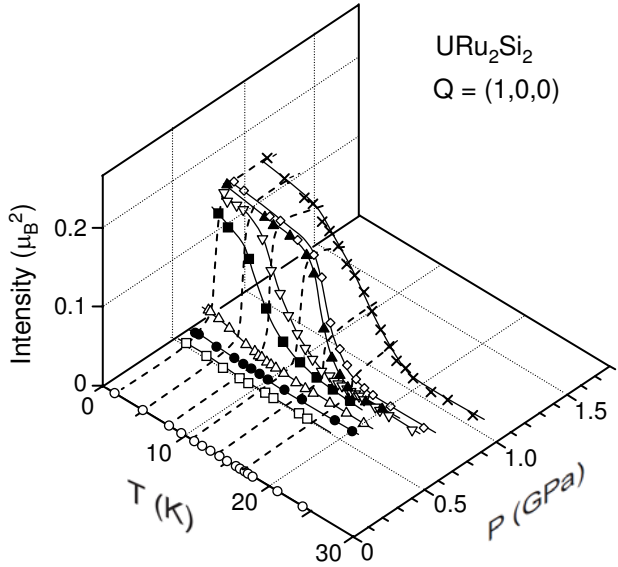}
\caption{
Overall view of the pressure-temperature dependence of the integrated magnetic Bragg-scattering intensity at $Q = (1,0,0)$ in {\urs}. The lines are guides to the eye. 
}
  	\label{fig-2}
  \end{figure}

  \begin{figure}[t]
\includegraphics[scale =1]{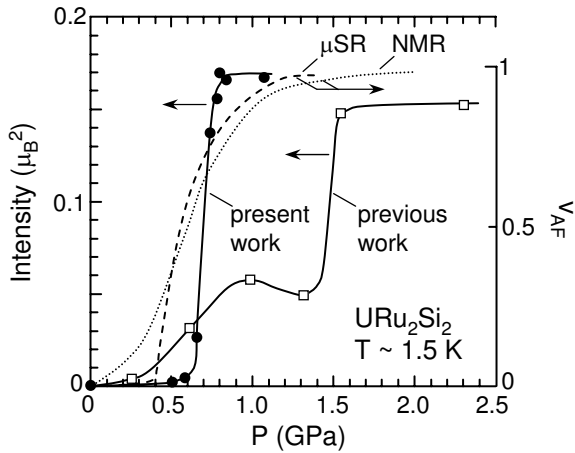}
\caption{
Pressure dependence of the integrated intensity of the (1,0,0) magnetic Bragg peak measured at 1.5 K for our present (closed circles) and previous (open squares \cite{ami99}) {\urs} single crystals. Pressure dependence of the AF volume fraction obtained from the previous $^{29}$Si-NMR \cite{matsuda01} and ${\mu}$SR \cite{ami02,amato04} measurements are also plotted on an arbitrary scale.
}
  	\label{fig-3}
  \end{figure}

The sharpening of the transition achieved now allow us to determine the transition point $P_{\rm M}(T)$ (or $T_{\rm M}(P)$)  from the $I_{\rm AF}(P)$ curves, with reasonable accuracy. Here, we define $P_{\rm M}$ as the midpoint of the increase in $I_{\rm AF}(P)$ at each constant temperature. The results are shown in Fig. 4 by open circles. The phase boundary line obtained starts at $P_{\rm M}(T \to 0) =$ 0.7 GPa, and runs upward in a very steep gradient. In the present study, however, the above definition of $P_{\rm M}$ can be adopted up to $\sim$ 16 K, because $I_{\rm AF}(P)$ does not reach the saturation for higher temperatures, within the present $P$ window. In addition, {\tord} and {\taf} cannot be detected by the present techniques, so that we plotted in Fig. 4 the data points obtained by electrical resistivity measurements using a different sample and pressure conditions. Therefore, we do not confirm the presence or absence of the bicritical point from the present study. Nevertheless, the significant decrease in $I_{\rm AF}$ for $P < P_{\rm M}$ concomitant with the sharpening of the $P$-induced transition strongly suggests that the HO breaks different symmetry from the AF order, i.e., the presence of the bicritical point.  

  \begin{figure}[t]
\includegraphics[scale =1]{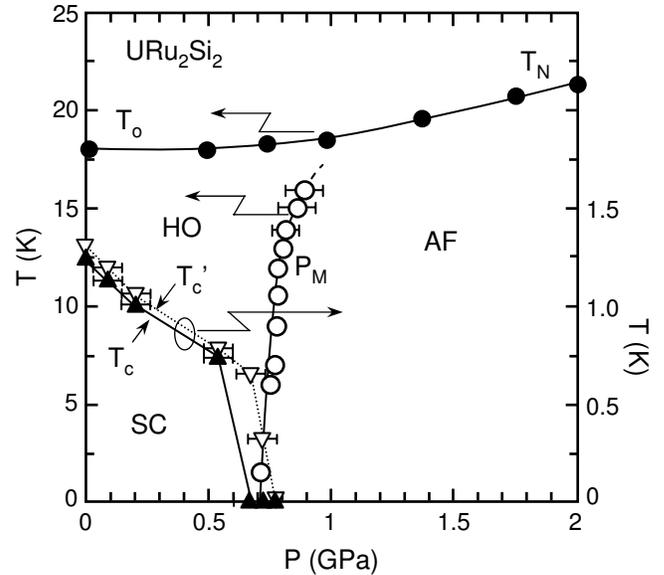}
\caption{
Pressure-temperature phase diagram of {\urs}. The phase boundary line between the HO and AF phases is formed by connecting the ``transition pressures'' $P_{\rm M}$ (open circles) determined from the present neutron-scattering measurements. Closed and open triangles represent the superconducting transition temperature {\tsc} defined as the temperature at which d${\chi}'$/d$T$ takes maximum, and the onset temperature {\tso} where ${\chi}'$ starts to decrease. The transition temperatures {\tord} and {\taf} (closed circles) are obtained from the electrical resistivity measurements using a different single crystal and a different high-pressure device. The lines are guides to the eye.
}
  	\label{fig-4}
  \end{figure}

  \begin{figure}[t]
\includegraphics[scale =1]{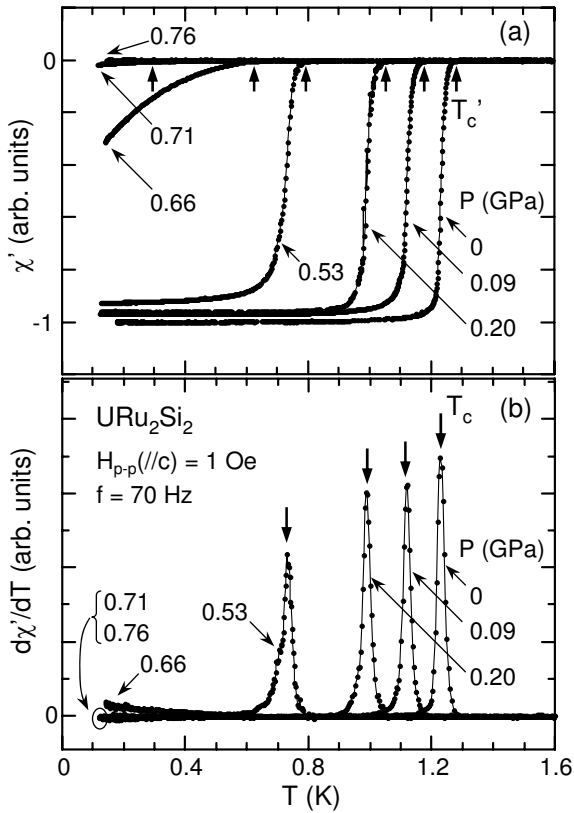}
\caption{
Temperature dependence of (a) the real part of ac susceptibility ${\chi}'$ and (b) its temperature derivative d${\chi}'$/d$T$ of {\urs}, measured at several applied pressure points. Backgrounds and the normal state contribution are subtracted.
}
  	\label{fig-5}
  \end{figure}

The superconductivity of the present sample also exhibits a dramatic change at $\sim$ 0.7 GPa. We define the SC transition temperature {\tsc} as the peak position of $d{\chi}'/dT$ and {\tso} as the onset of decrease in ${\chi}'$, where ${\chi}'$ denotes the real part of the ac susceptibility. The SC volume fraction $v_{\rm SC}$ was simply estimated by the magnitude of decrease in ${\chi}'$. As seen in Fig. 5, the SC transition is very sharp for $P \le$ 0.53 GPa, but becomes abruptly suppressed at $\sim$ 0.7 GPa. At 0.76 GPa, the transtion cannot be detected at least down to $\sim$ 0.1 K. Pressure dependences of {\tsc}, {\tso} and $v_{\rm SC}$ are summarized in Fig. 6. The experimental results clearly demonstrate that the sudden termination of SC phase is caused by the evolution of AF phase, and that they are not coexistent. Interestingly, the obtained $T_{\rm c}(P)$ curve agrees fairly well with that reported by Uemura et al. \cite{uemura05}, if they are normalized by $T_{\rm c}(0)$ and $P_{\rm M}(0)$ (the broken lines in Fig. 6). By contrast, the $v_{\rm SC}(P)$ curves disagree with each other. The gradual change of their $v_{\rm SC}$ data reflects a wider phase-separation range in their sample. These comparisons also confirm that the superconductivity occurs only in the HO region of the crystal. Considering the upper critical fields and the coherence lengths of this compound, the suppression of the superconductivity in the AF phase probably cannot be ascribed to the generation of the staggered internal fields. The development of the AF order may rather be linked to a complete suppression of the pairing interactions. It might be suggestive that the strong magnetic excitations in the HO state vanish in the high-$P$ AF phase, at least for $q = (1, 0, 0)$ and (1, 0.4, 0) \cite{ami99}. Because of the strong uniaxial magnetic anisotropy, the low-energy magnetic fluctuations of this system are limited to the longitudinal modes.  Therefore, once the moments are frozen, there will remain only high-energy magnetic fluctuations, which may not contribute to the pairing interactions.

  \begin{figure}[t]
\includegraphics[scale =1]{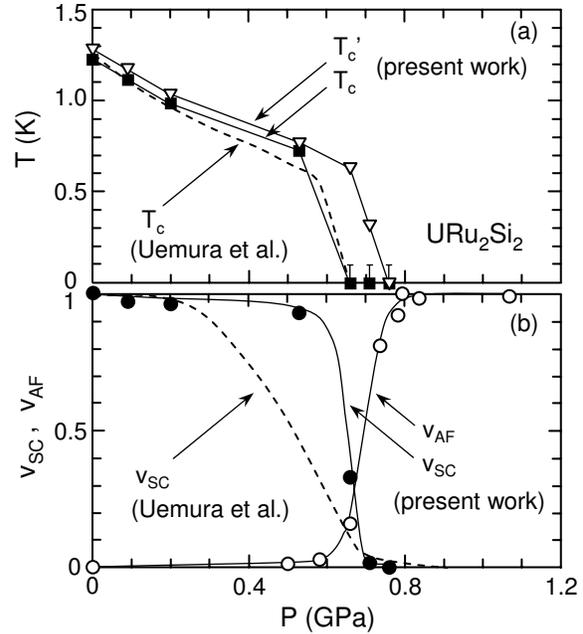}
\caption{
Pressure dependence of (a) the superconducting transition temperature {\tsc} and the onset temperature {\tso} and (b) the volume fractions $v_{\rm SC}$ and $v_{\rm AF}$ of the superconducting and antiferromagnetic phases of {\urs}. For comparison, the recent results of {\tsc} and $v_{\rm SC}$ reported by Uemura et al. \cite{uemura05} are also shown by nomalizing with $T_{\rm c}(0)$ and $P_{\rm M}(0)$ (the broken lines). The solid lines are guides to the eye.
}
  	\label{fig-6}
  \end{figure}

Finally, we briefly comment the FM behavior of {\urs}. In addition to the anomaly at $\sim$ 35 K ($\equiv T_{2}^{\star}$) pointed out by Uemura et al., we now show that the system may exhibit at least two other FM behavior below $T_{1}^{\star} =$ 120(5) K and $T_{3}^{\star} =$ 16.5(5) K (Fig. 7). We checked several batches of {\urs} single crystals and alloys U(Ru$_{1-x}$Rh$_{x}$)$_{2}$Si$_{2}$ ($x \le 0.04$), and observed that all the anomalies strongly depend on the sample quality. We also found from the NS measurements that at least the order developing below $T_{2}^{\star}$ is not FM but AF  with the same $Q = (1, 0, 0)$. Therefore, the FM behavior below $T_{2}^{\star}$ probably arises from the stacking faults of this AF phase. It is known that the Rh- doping suppresses both the HO and SC states and leads to a normal HE state at $x \sim$ 0.04 \cite{yokoyama04}. Also in the range $0.02 < x < 0.04$, the AF order appears similarly to the effects of hydrostatic pressure. In spite of such variations of the low-temperature phases, the three kinds of FM behavior remain nearly the same, except that their onset temperatures are slightly increasing (not shown). Although their origin remains open, we can thus safly say that the FM anomalies are extrinsic due to the existence of the residual impurity phases which may not be important to the intrinsic physical property of this compound.

  \begin{figure}[t]
\includegraphics[scale =1]{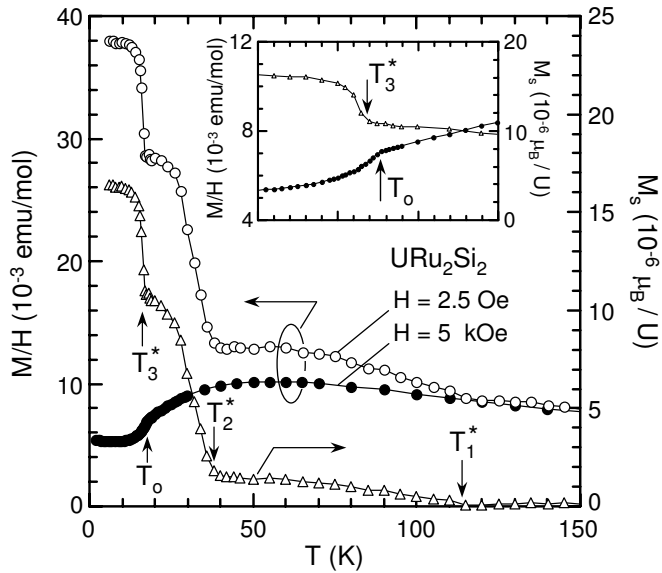}
\caption{
Temperature dependence of the linear magnetic susceptibility of {\urs} measured at 2.5 Oe (open circles) and 5 kOe (closed circles). Triangles show the spontaneous magnetization $M_{\rm s}$ at 2.5 Oe obtained from the deference between these two quantities. The inset shows an enlarged view around {\tord}.
}
  	\label{fig-7}
  \end{figure}

\section{Conclusions}
The neutron-scattering experiments using a thinner {\urs} single crystal have revealed that improving the quality of sample and pressure may significantly weaken the low-$P$ AF behavior and sharpen the $P$-induced transition between the HO and AF phases. This implies that the AF signal at the low-$P$ range is irrelevant to HO, and that HO breaks different symmetries from those for the high-$P$ AF order, suggesting the possibility of scenarios in category B2 given in Sec. 1. The so-called ``small-moment antiferromagnetic phase'' will be unlikely to be present as the uniform order. 

The ac-susceptibility measurements performed by using the same sample under the same pressure conditions have clearly shown that an abrupt and complete suppression of the SC phase occurs simultaneously with the development of the AF phase under high pressure. We thus conclude that the SC phase does not coexist with the AF phase in this system, supporting the proposal in Ref. 38. We have also pointed out that the {\urs} samples may contain, at least, three magnetic impurity phases of the order of 1-10 ppm in concentration. They exhibit very weak ferromagnetic behavior not only below $\sim$ 35 K, which was previously known, but also below $\sim$ 120 K and $\sim$ 16.5 K. Contrary to the recent suggestions \cite{uemura05}, the strong sample-quality dependence of the phenomena and the Rh-doping effects imply that all these phases do not essentially coupled to the HO, AF and SC phases. 

Since the present investigations clearly demonstrated that a volume fraction of the low temperature AF long range order phase below 0.7 GPa can be minimized, the intrinsic properties of HO phase, such as the superconductivity will be able to be clarified in near future. Further experimental studies are planning in our group.

\section*{Acknowledgments}
This work was supported, in part, by the Grant-in-Aids for the 21st Century COE ``Topological Science and Technology'' and for ``Evolution of New Quantum Phenomena Realized in the Filled Skutterudite Structure'' (No. 18027001) from the Ministry of Education, Culture, Sports, Science and Technology (MEXT) of Japan

\end{document}